%% file: dim_4-eps.tex
\documentclass[amssymb, nofootinbib, superscriptaddress]{revtex4}

\input{sections/preamble}

\begin{document}



\title{Group Field Theory in dimension $\bf 4 - \varepsilon $}

\author{Sylvain Carrozza}\email{sylvain.carrozza@cpt.univ-mrs.fr}\affiliation{Centre de Physique Th\'eorique\\ 
CNRS UMR7332, Universit\'e d'Aix-Marseille, 13288 Marseille cedex 9, France}

\begin{abstract}
\bigskip
Building on an analogy with ordinary scalar field theories, an $\varepsilon$-expansion for rank-$3$ tensorial group field theories with gauge invariance condition is introduced. This allows to continuously interpolate between the dimension four group $\SU(2) \times \U(1)$ and the dimension three $\SU(2)$. In the first situation, there is a unique marginal $\vphi^4$ coupling constant, but in contrast to ordinary scalar field theory this model is asymptotically free. In the $\SU(2)$ case, the presence of two marginally relevant $\varphi^6$ coupling constants and one $\varphi^4$ super-renormalizable interaction spoils this interesting property. However, the existence of a non-trivial fixed point is established in dimension $4 - \varepsilon$, hence suggesting that the $\SU(2)$ theory might be asymptotically safe. To pave the way to future non-perturbative calculations, the present perturbative results are discussed in the framework of the effective average action. 
\end{abstract}

\maketitle


\section{Introduction}
\input{sections/intro}

\section{Rank-$3$ tensorial group field theory with gauge invariance}\label{sec:tgfts}
\input{sections/background}

\section{Asymptotically free model on $\SU(2) \times \U(1)$}\label{sec:4d}
\input{sections/dim4}

\section{Small perturbation of the group dimension}\label{sec:4d-e}
\input{sections/dim4_e}

\section{Towards a UV completion of the $\SU(2)$ model}\label{sec:3d}
\input{sections/su2}

\section{Conclusion}
\input{sections/conclu}




\bibliographystyle{hunsrt}
\bibliography{biblio}

\end{document}

%% file: sections/preamble.tex
\usepackage{hyperref}
\usepackage{amsfonts,amssymb,amsmath,exscale,bbm}
\usepackage{footnpag} 
\usepackage[all,knot]{xy}  
\usepackage{color}
\usepackage{graphicx}
\usepackage{epsfig}
\usepackage{subfig}
\usepackage{enumerate}

\usepackage{bbold}

\usepackage{amsmath}
\usepackage{amsfonts}
\usepackage{graphicx}
\usepackage{psfrag}
\usepackage{array}
\usepackage{bbm}
\usepackage{hyperref}
\usepackage{amsthm}
\theoremstyle{definition}

\newcommand{\cN}{{\mathcal N}}
\newcommand{\cO}{{\mathcal O}}

\newcommand{\cR}{{\mathcal R}}

\newcommand{\cZ}{{\mathcal Z}}

\def\inv{{\mbox{\tiny -1}}}

\newcommand\beq{\begin{equation}}
\newcommand\eeq{\end{equation}}
\newcommand{\be}{\begin{equation}}
\newcommand{\ee}{\end{equation}}
\newcommand{\bes}{\begin{eqnarray}}
\newcommand{\ees}{\end{eqnarray}}
\newcommand{\bea}{\begin{eqnarray}}
\newcommand{\eea}{\end{eqnarray}}

\def\vphi{{\varphi}}
\def\vphib{\overline{{\varphi}}}

\def\tg{\tilde{g}}

\newcommand{\one}{\mbox{$1 \hspace{-1.0mm}  {\bf l}$}}

      \def\nn{{\nonumber}}

\newcommand{\SU}{\mathrm{SU}}

\newcommand{\U}{\mathrm{U}}

\def\extd{\mathrm {d}}

\newcommand\acts\triangleright

\newcounter{letter} \newcounter{numeral} \newcounter{Numeral}

\def\vphi{\varphi}

\def\extd{\mathrm {d}}



%% file: sections/intro.tex






Group Field Theory (GFT) \cite{freidel_gft, daniele_rev2006, daniele_rev2011, Krajewski_rev} is a general formalism aiming at completing the definition of the dynamics of Loop Quantum Gravity (LQG) \cite{ashtekar_book, rovelli_book, thiemann_book, bojowald_book, rovelli_vidotto}, either from a covariant perspective as was historically proposed and since then has been the main line of investigation \cite{dPFKR, GFT_rovelli_reisenberg}, or directly from the canonical picture as was more recently suggested \cite{daniele_2nd, gft_lqg_2}. 
An alternative but related approach to the same question relies on lattice gauge theory methods \cite{bianca_cyl, bianca_sebastian, bianca_continuum_2014, bahr2014}. In both Wilson's renormalization group is central, first to consistently define the theory, and at a later stage to explore its phase structure. In the long run, we hope to understand the effective, low energy limit of LQG, and be in a position to check whether Einstein's gravity is reproduced or not.

Mathematically speaking, GFTs are quantum field theories defined on group manifolds, and with a peculiar type of combinatorially non-local interactions. Thanks to recent breakthroughs in the closely related field of tensor models \cite{razvan_jimmy_rev, tt2, tt3}, the usual quantum field theory tools are currently being generalized to GFTs which increasingly resemble LQG. These developments include first and foremost perturbative renormalization \cite{tensor_4d, josephaf, josephsamary, cor_u1, cor_su2, samary_vignes, joseph_etera, samary_beta}, but also constructive results \cite{razvan_beyond, delepouve_borel, samary_2point}, and generalizations of the usual functional renormalization group (FRG) methods \cite{polch_gft, frgft}.

\

In this paper we continue our exploration of tensorial GFTs (TGFTs) with gauge invariant condition \cite{cor_u1, cor_su2, thesis}. These are GFTs combinatorially inspired by tensor models \cite{universality, uncoloring}, but with a loop quantum gravity flavour at the group--theoretic level: the so--called closure constraint, present in all spin foam models \cite{perez_review2012, Baez:1997zt, oriti2001spacetime, Alexandrov:2011ab}, is implemented. Among TGFTs already known to be perturbatively renormalizable, the rank-$3$ $\SU(2)$ model first introduced in \cite{cor_su2} is arguably the closest to loop quantum gravity: technically because its boundary states are $\SU(2)$ spin networks; and conceptually because, except for its Laplace-type propagator, all its other ingredients can be understood as arising from the GFT quantization of peculiar cellular discretizations of 3d Euclidean gravity. 

In \cite{discrete_rg}, a discrete version of Wilson's renormalization was applied to the study of the Gaussian fixed point of this model. It was proven that, despite the domination of wave--function renormalization over vertex renormalization, positive perturbations of its two $\vphi^6$ interactions are incompatible with asymptotic freedom. A natural question to investigate is therefore that of the existence of non-trivial fixed points, which might provide a natural UV completion of this model. This is a difficult task, which can only be fully understood with non-perturbative methods such as the FRG. In this paper, we will stay at the perturbative level, but will invoke an $\varepsilon$-expansion similar to \cite{wilson_fisher, wilson_kogut_epislon} to forecast the qualitative properties of the $\SU(2)$ model. As we will argue below, this might be interesting \emph{per se}. But in the spirit of our previous works on this topic, we primarily view this type of study as an opportunity to develop the GFT framework further, and thus make new technical tools available for future investigations in the physical four--dimensional context.

\

The plan of the paper is as follows. Section \ref{sec:tgfts} introduces the necessary background on rank-$3$ TGFTs with gauge invariance condition. In section \ref{sec:4d} we focus on a $\vphi^4$ renormalizable model on the four--dimenional group $\SU(2) \times \U(1)$. Using the language of the effective average action \cite{wetterich_eq, wetterich_2001}\footnote{See also \cite{review_ren_GRS} for an introduction to FRG methods, among other advanced aspects of renormalization.}, we prove it asymptotically free. In section \ref{sec:4d-e} we analytically continue the dimension of the $\U(1)$ factor so as to formally define the theory on a group of dimension $4 - \varepsilon$. Just like in ordinary scalar field theory on a Euclidean space of dimension $4- \varepsilon$, we prove the existence of a non--trivial fixed point at a distance of order $\varepsilon$ from the origin, with one relevant direction less than the Gaussian fixed point. Finally, in section \ref{sec:3d}, we recast some of the results of \cite{discrete_rg} about the $\SU(2)$ model in the continuous language of the effective average action, and we provide some perspectives for the future. Note that we will spare the reader with unnecessary technical details, all of which can be found in more or less ready-to-use form in \cite{cor_su2, thesis, discrete_rg}. 

%% file: sections/background.tex
\subsection{Definitions}

In order to introduce the type of models we will be concerned with in the following, let $G$ be a compact Lie group of dimension $D$. A rank-$3$ complex GFT over this group is a quantum field theory for a (complex) field $\vphi( g_1 , g_2 , g_3 )$, with $g_\ell \in G$. 

\begin{figure}[h]
  \centering
 	\includegraphics[scale=0.6]{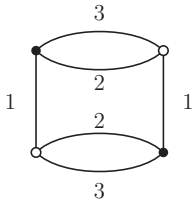}     
  \caption{A four-valent bubble $b$.}\label{ex_bubble}
\end{figure}

The tensoriality criterion is a restriction on the type of interactions we allow in this theory. We require the interaction part of the action $S^{{\rm{int}}}$ to be a sum of \emph{connected tensor invariants}. They are given by specific convolutions of the elementary fields $\vphi$ and $\vphib$ which are in one-to-one correspondence with \emph{connected $3$-colored graphs}. Such a graph is made out of two types of nodes (black or white), and three types of edges (of color $\ell \in \{ 1 \,, 2 \,, 3 \}$), with the following rules: a) only nodes of different types can be connected by an edge (i.e. the graph is bipartite); b) to each node is hooked exactly one edge of each color. Such graphs are also called \emph{bubbles}. Finally, to each bubble $b$ is associated a unique tensorial invariant $I_b (\vphi , \vphib )$ as follows: each white (resp. black) node corresponds to a field $\vphi$ (resp. $\vphib$); and an edge of color $\ell$ between two fields is understood as a convolution of their $\ell^{\rm{th}}$ arguments. As an example, the invariant associated to the bubble $b$ represented in Figure \ref{ex_bubble} is:
\beq
I_b ( \vphi , \vphib ) = \int [\extd g_i]^6 \, \vphib (g_6 , g_2 , g_3 ) \vphi (g_1 , g_2 , g_3 ) \vphib (g_1 , g_4 , g_5 )  \vphi (g_6 , g_4 , g_5 )\,,  
\eeq
where $\extd g_i$ is the Haar measure associated to the variable $g_i$.

\

The action $S$ is then assumed to be of the form:
\beq
S ( \vphi , \vphib ) = S_\vphi ( \vphi , \vphib ) + S^{{\rm{int}}} ( \vphi , \vphib )\,,
\eeq
where 
\begin{align}
S_\vphi ( \vphi , \vphib ) &= - \int [\extd g_\ell ]^3 \, \vphib( g_1 , g_2 , g_3 ) \left( \sum_{\ell=1}^3 \Delta_\ell \right) \vphi( g_1 , g_2 , g_3 )\,, \\
S^{{\rm{int}}} ( \vphi , \vphib ) &= \sum_b t_b I_b ( \vphi , \vphib )\,,  \label{action_dim}
\end{align}
and $\Delta_\ell$ is the Laplace--Beltrami operator acting on the $\ell^{\rm{th}}$ variable of the field. The coupling constants $t_b$ are more over chosen so as to guarantee the invariance of the theory under arbitrary permutations of the color labels\footnote{This is not necessary as far as the consistency of the field theory is concerned, but is very natural from the point of view of discrete geometry and gravity.}. Note that we included the mass term in $S^{{\rm{int}}}$.

\

The last ingredient is the \emph{gauge invariance condition}. It is a symmetry of the field under simultaneous translation of its arguments:
\beq\label{gauge_inv}
\forall h \in G \, , \qquad \vphi( g_1 , g_2 , g_3 ) = \vphi( g_1 h, g_2 h, g_3 h)\,,  
\eeq
which at the level of the amplitudes introduces a discrete $G$-connection, hence its (somewhat misleading) name. Because there exists no well--defined Lebesgue measure on the space of invariant fields, this constraint is combined with the kinetic action $S_\vphi$ to yield a (degenerate) Gaussian measure on the space of non--invariant fields. The theory is thus formally defined by the partition function:
\beq
\cZ = \int \extd \mu_C ( \vphi, \vphib ) \exp\left( - S^{{\rm{int}}} ( \vphi , \vphib ) \right)\,, 
\eeq    
where $\mu_C$ is the Gaussian measure of covariance:
\beq
C(g_1 , g_2 , g_3 ; \tg_1 , \tg_2 , \tg_3 ) =  \int_{0}^{+ \infty} \extd \alpha  \int \extd h \, \prod_{\ell = 1}^3 K_{\alpha} ( g_\ell h \tg_\ell^\inv)\,,
\eeq
and $K_\alpha$ is the heat-kernel at time $\alpha$ on $G$. This is the usual Schwinger representation of the propagator, with an additional gauge averaging over the group which projects onto the space of invariant fields (\ref{gauge_inv}).  

\subsection{Renormalizability and canonical dimensions}

A general power--counting theorem for TGFTs with gauge invariance condition was first derived in \cite{cor_u1, cor_su2}, with the help of older results \cite{lin_gft}. A classification of potentially renormalizable models could thus be deduced, in terms of the rank $d$ of the fields, the dimension $D$ of the group and the maximal valency of renormalizable interactions $v_{max}$. For $d=3$, there are exactly two possibilities:
\begin{itemize}
\item a $\vphi^6$ model in $D=3$;
\item a $\vphi^4$ model in $D=4$.
\end{itemize}
This is very reminiscent of scalar field theories on Euclidean space-times, for which exactly the same thing occurs provided that $D$ is understood as the space-time dimension. 

\

In \cite{discrete_rg}, appropriate scalings of the coupling constants $t_b$ in equation (\ref{action_dim}) were deduced from the power--counting theorem. In order to systematically take them into account we introduced the notion of \emph{canonical dimension} $d_b$ of a coupling constant $t_b$, which when $d=3$ is defined as:
\beq\label{can_dim}
d_b = [t_b] = D  -  \frac{ (D - 2) N_b}{2}\,. 
\eeq
Just like in ordinary space-time based field theories, this dimension provides a simple criterion to determine the relevant directions in the vicinity of the Gaussian fixed point: renormalizable coupling constants must have positive or zero canonical dimensions, while non-renormalizable interactions come with a strictly negative canonical dimension. In the Wilsonian approach, it is important to introduce dimensionless parameters:
\beq
t_b  = u_b \mu^{d_b}\,, 
\eeq
where $\mu$ is an appropriately chosen scale. In the following sections, this role will be played by the infrared floating cut-off scale of the effective average action.

%% file: sections/dim4.tex
%

\subsection{The model and its $\beta$-functions in the ultraviolet region}

In this section we consider the rank-$3$ TGFT based on the group $G = \SU(2) \times \U(1)$. In addition to the mass, there is only one marginally relevant $\vphi^4$ coupling constant on such a dimension $4$ group. We therefore parameterize the bare theory at cut-off scale $\Lambda$ by:
 \begin{align}
 \cZ_{\Lambda} &= \int \extd \mu_{C_{\Lambda}} (\vphib , \vphi) \exp \left( - S_\Lambda^{{\rm{int}}} (\vphib , \vphi) \right) \,, \\
 S_\Lambda^{{\rm{int}}} (\vphib , \vphi) &= \Lambda^2 u_{2, \Lambda} S_{2}(\vphi , \vphib) + \frac{u_{4 , \Lambda}}{2} S_4 (\vphib , \vphi) \,,\\
 C_{\Lambda} (g_1 , g_2 , g_3 ; \tg_1 , \tg_2 , \tg_3 )&=  \int_{\Lambda^{-2}}^{+ \infty} \extd \alpha  \int \extd h \, \prod_{\ell = 1}^3 K_{\alpha} ( g_\ell h \tg_\ell^\inv)\,,
 \end{align}
where $S_2$ contains the mass term and $S_4$ is the sum of all $\vphi^4$ interactions (that is three bubbles of the type shown in Figure (\ref{ex_bubble})):
\begin{align}
S_{2} (\vphi , \vphib) &= \int [\extd g ]^3 \, \vphi(g_1 , g_2 , g_3 ) \vphib(g_1 , g_2 , g_3 )\,,\\
S_{4} (\vphi , \vphib) &= \int [\extd g ]^6 \, \vphi(g_1 , g_2 , g_3 ) \vphib(g_1 , g_2 , g_4 ) \vphi(g_5 , g_6 , g_3 ) \vphib(g_5 , g_6 , g_4 ) + \; {\rm two} \; {\rm color} \; {\rm permutations}  \,.
\end{align}

\

We now introduce our renormalization scheme, based on the effective average action. In this approach the UV cut--off can be assumed large and fixed, and will eventually be removed. The scale parameterizing the flow is an infrared scale $k$, which we will assume small as compared to $\Lambda$. But since we are interested in the UV behavior of the theory, we must also require $k$ to be large, and hence we assume:
\beq\label{scales}
1 \ll k \ll \Lambda\,.
\eeq
We then introduce a functional $\widetilde{\Gamma}_k$, implicitly defined by:
\beq\label{gammat}
\exp\left( {- \widetilde{\Gamma}_k} (\vphib , \vphi) \right) = \frac{1}{\cN_\Lambda^k} \int \extd \mu_{C_{\Lambda}^k} (\overline{\phi}, \phi) \exp \left( - S_\Lambda^{{\rm{int}}} (\vphib + \overline{\phi} , \vphi + \phi ) + \overline{\phi} \cdot \frac{\delta \widetilde{\Gamma}_k}{\delta \overline{\vphi}} (\vphib , \vphi) + \phi \cdot \frac{\delta \widetilde{\Gamma}_k}{\delta \vphi} (\vphib , \vphi)  \right)
\eeq
where
\begin{align}
 C_{\Lambda}^k (g_1 , g_2 , g_3 ; \tg_1 , \tg_2 , \tg_3 ) &=  \int_{\Lambda^{-2}}^{k^{-2}} \extd \alpha  \int \extd h \, \prod_{\ell = 1}^3 K_{\alpha} ( g_\ell h \tg_\ell^\inv)\,, \\
 \cN_\Lambda^k &= \int \extd \mu_{C_{\Lambda}^k} (\overline{\phi}, \phi) \exp \left( - \Lambda^2 u_{2, \Lambda} S_{2}(\vphi , \vphib) \right)\,.
\end{align}
This type of implicit definition is quite appropriate in our context because it does not require the introduction of a formal Lebesgue measure on the space of invariant fields, but rather makes direct reference to the well--defined Gaussian measure $\extd \mu_{C_{\Lambda}^k}$. We however need to reintroduce the kinetic part of the action a posteriori to obtain the \emph{effective average action} at scale $k$:
\beq
\Gamma_k \equiv S_\vphi + \widetilde{\Gamma}_k\,.   
\eeq
It interpolates between the bare total action
\beq
\Gamma_\Lambda = S_\vphi + S_\Lambda^{{\rm{int}}} = S_\Lambda
\eeq
and the full effective action $\Gamma_0$. 

At the perturbative level, $\Gamma_k$ can be computed order by order in a loop expansion, with only connected one particle irreducible graphs contributing. In this paper, we will only consider one-loop corrections, which amounts to expanding the action to second order in the fluctuating fields $\phi$ and $\overline{\phi}$ in the right-hand-side of equation (\ref{gammat}).
We will use the following parameterization
\beq
\Gamma_k (\vphib , \vphi) = {Z_k}  k^2 u_{2,k}   S_2 (\vphib , \vphi) + {Z_k} S_\vphi (\vphib , \vphi) + {Z_k}^2 \frac{u_{4 , k}}{2} S_4 (\vphib , \vphi)  + \cR_{\Lambda}^k (\vphib , \vphi) \,,
 \eeq
 where $\cR_{\Lambda}^k (\vphib , \vphi)$ is a sum of convergent graphs, finite Taylor remainders and higher loop contributions. $u_{2,k}$ and $u_{4,k}$ are the one-loop dimensionless coupling constants at cut-off scale $k$ and $Z_k$ encodes the wave-function renormalization. In the following we will moreover denote by $\cO({u_\Lambda}^n)$ all the terms which are $\cO({u_{2,\Lambda}}^k {u_{4,\Lambda}}^{n-k})$ for some $k \in \{ 0  , \ldots , n\}$. One finds that:
 \begin{align}
 {Z_k}^2 k^2 u_{2, k} &= \Lambda^2 u_{2,\Lambda} + 3 a_{k , \Lambda}  u_{4, \Lambda} + \cO({u_{\Lambda}}^2) \,, \label{eq2}\\
 {Z_k}^2 u_{4, k} &= u_{4, \Lambda} - b_{k , \Lambda} {u_{4, \Lambda}}^2 + \cO({u_{\Lambda}}^3)\,, \label{eq4}\\
 Z_k &= 1 - w_{k , \Lambda} u_{4 , \Lambda} + \cO({u_{\Lambda}}^2) \,,
 \end{align}
 where $a_{k , \Lambda}$ and $b_{k , \Lambda}$ are vertex corrections, respectively generated by the graphs $G_1^\ell$ and $G_2^\ell$, while $w_{k , \Lambda}$ is a wave-function correction due to the second term in the Taylor expansion of the graph $G_1^\ell$. 

 \begin{figure}[h]
  \centering
  \subfloat[$G_{1}^\ell$]{\label{g1l}\includegraphics[scale=0.6]{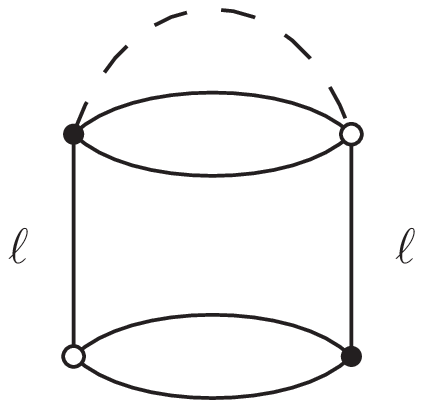}}     
  \subfloat[$G_{2}^{\ell}$]{\label{g2l}\includegraphics[scale=0.6]{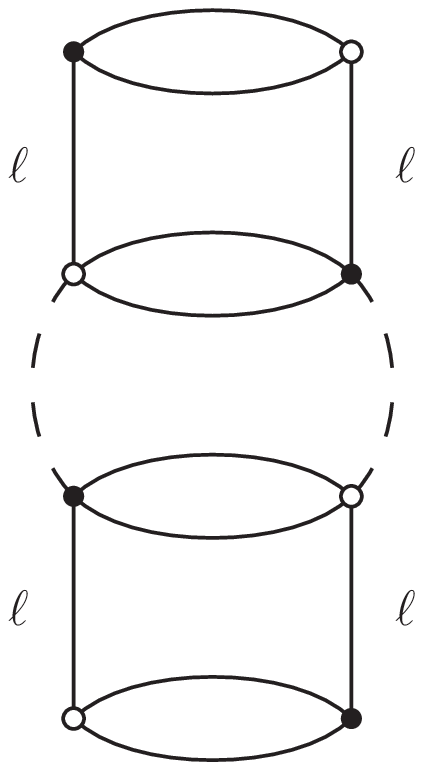}}
  \caption{First graphs contributing to the flow equations: the dashed lines represent propagators.}\label{2graphs}
\end{figure}

We refer to \cite{discrete_rg} for details about the combinatorial factors entering this expansion and for a description of how such amplitudes should be computed (also \cite{cor_su2, thesis} are valuable in this respect). One can show that
  \begin{align}
 a_{k , \Lambda} &\approx \int_{\Lambda^{-2}}^{k^{-2}} \extd \alpha \int \extd h \, [K_{\alpha} (h)]^2  \,, \label{a2} \\
  b_{k , \Lambda} &\approx \iint_{\Lambda^{-2}}^{k^{-2}} \extd \alpha_1 \extd \alpha_2 \iint \extd h_1 \extd h_2 \, [K_{\alpha_1 + \alpha_2} (h_1 h_2)]^2 \,, \label{a4}\\
 w_{k , \Lambda} \Delta_G &\approx  \frac{1}{2!} \int \extd g \left( \frac{\vert X_g\vert^2}{3} \Delta_{\SU(2)} + \vert \theta_g \vert^2 \Delta_{\U(1)} \right) \int_{\Lambda^{-2}}^{k^{-2}} \extd \alpha \int \extd h \, [K_\alpha (h)]^2 K_{\alpha} (hg)\,, \label{w}
 \end{align}
 where $X_g \in \SU(2)$ is the logarithm of the $\SU(2)$ factor of $g$, and similarly $\theta_g$ represents the angle coordinates of the $\U(1)$ factor. The sign $\approx$ indicates that the short-time asymptotics of the heat-kernel is invoked to evaluate the right-hand-sides. It is only in this limit, justified by our assumption (\ref{scales}), that we will obtain an autonomous system of flow equations (see \cite{frgft} for a general perspective on the non--autonomous character of the renormalization group flow in GFTs).

 Solving for the wave-function normalization $Z_k$ in (\ref{eq4}) and (\ref{eq2}) we obtain:
  \begin{align}
 k^2 u_{2, k} &= \Lambda^2 u_{2,\Lambda} + 3 a_{k , \Lambda}  u_{4, \Lambda} + \cO({u_{\Lambda}}^2) \,,\\
 u_{4, k} &= u_{4, \Lambda} + \left( 2 w_{k , \Lambda} - b_{k , \Lambda} \right) {u_{4, \Lambda}}^2 + \cO({u_{\Lambda}}^3)\,, 
 \end{align}
 which shows that the qualitative evolution of the $4$-valent coupling constant $u_4$ will be the result of a competition between the vertex and wave-function renormalizations. Acting with the operator $k \frac{\partial}{\partial k}$ on these two equations, and then reexpressing the constants at scale $\Lambda$ in terms of those at scale $k$ at tree level, we finally obtain:
 \begin{align}
 \beta_2 (u_{2,k}, u_{4,k}) &\equiv k \frac{\partial u_{2,k}}{\partial k} = - 2 u_{2,k} + \frac{3}{k} \frac{\partial a_{k , \Lambda}}{\partial k}  u_{4, k} + \cO({u_{k}}^2)\,, \\ 
 \beta_4 (u_{2,k}, u_{4,k}) &\equiv k \frac{\partial u_{4,k}}{\partial k} =  \left( 2 k \frac{\partial w_{k , \Lambda}}{\partial k} - k \frac{\partial b_{k , \Lambda} }{\partial k} \right) {u_{4, k}}^2 + \cO({u_{k}}^3)\,. 
 \end{align}
Note that the first term in $\beta_2$ reflects the canonical dimension of $u_2$.
 
\subsection{Estimation of the $\beta$-functions}

We are now ready to evaluate the $\beta$-functions.
To this effect, first notice that the heat-kernel $K_\alpha$ on $G$ factorizes as a product
\begin{align}
\forall g = (h_1, h_2) \in \SU(2) \times \U(1) \,, \qquad K_\alpha (g) = K_\alpha^{\SU(2)}(h_1) K_\alpha^{\U(1)}(h_2) \,,
\end{align}
where $K_\alpha^{\SU(2)}$ (resp. $K_\alpha^{\U(1)}$) is the heat-kernel on $\SU(2)$ (resp. $\U(1)$).
The scale $k$ being assumed to be large, we can evaluate the expressions (\ref{a4}), (\ref{a2}) and (\ref{w}) by means of a Laplace approximation. This consists in the replacements:
\beq
\begin{split}
\int_{\SU(2)} \extd h_1^{(i)} \; &\rightarrow \; \frac{1}{16 \pi^2} \int_{\mathbb{R}^3}  \extd X^{(i)} \,, \\
\int_{\U(1)} \extd h_2^{(i)} \; &\rightarrow \; \frac{1}{2 \pi} \int_{\mathbb{R}} \extd \theta^{(i)} \,, 
\end{split}
\qquad
\begin{split}
K_\alpha^{\SU(2)} ( h_1^{(1)} \cdots  h_1^{(n)} ) \; &\rightarrow \; \frac{2\sqrt{\pi}}{\alpha^{3/2}} \exp\left( - \frac{\vert \sum_{i=1}^n X^{(i)}\vert^2}{4 \alpha} \right) \,,\\
K_\alpha^{\U(1)} ( h_2^{(1)} \cdots  h_2^{(n)} ) \; &\rightarrow \; \frac{\sqrt{\pi}}{\alpha^{1/2}} \exp\left( - \frac{\vert \sum_{i=1}^n  \theta^{(i)}\vert^2}{4 \alpha} \right) \,.
\end{split}
\eeq
in integrals which are sharply peaked around the configuration $\{ g^{(i)} = (h_1^{(i)} , h_2^{(i)} ) = \one \}$.
This procedure allows us to trade the group integrals for Gaussian integrals on Euclidean spaces. These can then be straightforwardly computed, yielding: 
 \beq\label{abw}
\begin{split}
 a_{k , \Lambda} &= \frac{\pi}{2} \int_{\Lambda^{-2}}^{k^{-2}} \frac{\extd \alpha}{\alpha^2} \,,\\
 b_{k , \Lambda} &= \frac{\pi}{2} \iint_{\Lambda^{-2}}^{k^{-2}} \frac{\extd \alpha_1 \extd \alpha_2}{(\alpha_1 + \alpha_2)^2 } \,, \\
 w_{k , \Lambda} &=  \frac{3 \pi}{4} \int_{\Lambda^{-2}}^{k^{-2}} \frac{\extd \alpha}{\alpha} \,,
\end{split}
\qquad \qquad 
\begin{split}
 \frac{1}{k} \frac{\partial a_{k , \Lambda}}{\partial k} &= - \pi \,,\\
 k \frac{\partial b_{k , \Lambda}}{\partial k} &= - \pi \frac{1 - \frac{k^2}{\Lambda^2}}{1 + \frac{k^2}{\Lambda^2}} \approx - \pi\,, \\
 k \frac{\partial w_{k , \Lambda}}{\partial k} &= - \frac{3 \pi}{2}\,.
\end{split}
 \eeq
 It thus turns out that wave-function terms dominate over vertex contributions in $\beta_4$, making it negative in the perturbative regime:
 \beq
 \left\{ 
 \begin{aligned}
 \beta_2 (u_{2}, u_{4}) &= - 2 u_2 - 3 \pi  u_{4} + \cO({u}^2)\,, \\
 \beta_4 (u_{2}, u_{4})  &=  - 2 \pi {u_{4}}^2 + \cO({u}^3)\,.
 \end{aligned}\right.
 \eeq
 Hence the Gaussian fixed point is UV-stable, and this model is asymptotically free. 
 In particular, a perturbative solution for $u_{4,k}$ can be labeled by a dynamically generated infrared scale $\Lambda_0$, such that: 
 \beq
 \forall k \gg \Lambda_0 \,, \qquad u_{4,k} = \frac{1}{2 \pi \ln\left( \frac{k}{\Lambda_{0}}\right)}\,.
 \eeq
 $\Lambda_0$ is analogous to the QCD scale $\Lambda_{\rm{QCD}}$: it is the scale around which one expects the perturbative treatment to break down and a phase transition to occur.

%% file: sections/dim4_e.tex
We now turn to the main purpose of this publication, which is to define an $\varepsilon$-interpolating between $D=3$ and $D= 4$.

\subsection{Analytic continuation to $4 - \varepsilon$ group dimension}

The analytic continuation of the group dimension is introduced via the $\U(1)$ factor of the group $\SU(2) \times \U(1)$. The idea is to first generalize the model of the previous section to a gauge group $G^{(D)} = \SU(2) \times \U(1)^{D}$, for an arbitrary $D \in \mathbb{N}$, and then analytically continue $D$ to $1 - \varepsilon$, with $\varepsilon > 0$ infinitesimally small. 

\

For instance, the quantities (\ref{abw}) extracted from the two graphs represented in Figure \ref{2graphs} become: 
\begin{align}
a_{k , \Lambda}^{(D)} &= \left(\frac{\pi}{2}\right)^{D/2 - 1} \int_{\Lambda^{-2}}^{k^{-2}} \frac{\extd \alpha}{\alpha^{D/2}} \,,\\
 b_{k , \Lambda}^{(D)} &= \left(\frac{\pi}{2}\right)^{D/2 - 1} \iint_{\Lambda^{-2}}^{k^{-2}} \frac{\extd \alpha_1 \extd \alpha_2}{(\alpha_1 + \alpha_2)^{D/2} } \,, \\
 w_{k , \Lambda}^{(D)} &=  \frac{3}{2} \left(\frac{\pi}{2}\right)^{D/2 - 1} \int_{\Lambda^{-2}}^{k^{-2}} \extd \alpha \, \alpha^{1 - D/2} \,.
\end{align}
These are all we need to compute the one-loop flow.

\subsection{Existence of a non-trivial fixed point}

When $D= 4 - \varepsilon$, the $4$-valent interactions acquire a non-zero canonical dimension:
\beq
d_4^{(4-\varepsilon)} = \varepsilon \,,
\eeq
while the dimension of the mass term is left unchanged.
We therefore parameterize the effective average action by:
\beq
\Gamma_k^{(4-\varepsilon)} (\vphib , \vphi) = Z_k  k^2 u_{2,k}   S_2 (\vphib , \vphi) + Z_k S_\vphi (\vphib , \vphi) + {Z_k}^2 k^\varepsilon \frac{u_{4 , k}}{2} S_4 (\vphib , \vphi)  + \cR_{\Lambda}^k (\vphib , \vphi) \,.
\eeq
Repeating the construction of the previous section, we easily compute the one-loop $\beta$-functions:
 \begin{align}
 \beta_2^{(4- \varepsilon)} (u_{2,k}, u_{4,k}) &= - 2 u_{2,k} + 3 k^{\varepsilon - 1} \frac{\partial a_{k , \Lambda}^{(4- \varepsilon)}}{\partial k}  u_{4, k} + \cO({u_{k}}^2)\,, \\ 
 \beta_4^{(4- \varepsilon)} (u_{2,k}, u_{4,k}) &=  - \varepsilon u_{4,k} + \left( 2 k^{1 + \varepsilon} \frac{\partial w_{k , \Lambda}^{(4- \varepsilon)}}{\partial k} - k^{1 + \varepsilon} \frac{\partial b_{k , \Lambda}^{(4- \varepsilon)} }{\partial k} \right) {u_{4, k}}^2 + \cO({u_{k}}^3)\,,
 \end{align}
 where the first term in the second line is a direct consequence of the new canonical dimension of $u_4$. The different coefficients entering these equations are easily computed:
 \bes
 k^{\varepsilon - 1} \frac{\partial a_{k , \Lambda}^{(4 - \varepsilon)}}{\partial k} &=& - \pi^{1 - \varepsilon} \,,\\
 k^{1 + \varepsilon} \frac{\partial b_{k , \Lambda}^{(4 - \varepsilon)}}{\partial k} &=& - \frac{4}{1 - \varepsilon / 2 } \left(\frac{\pi}{2}\right)^{1 - \varepsilon / 2} \left( \frac{1}{\left( 1 + (\frac{k}{\Lambda})^2 \right)^{1 - \varepsilon / 2}} - \frac{1}{2^{1 - \varepsilon / 2}} \right) \,, \\
 k^{1 + \varepsilon}  \frac{\partial w_{k , \Lambda}}{\partial k} &=& - 3 \left(\frac{ \pi}{2}\right)^{1 - \varepsilon / 2}\,. 
 \ees
 Thus expanding in $\varepsilon$ one finds:
\beq\label{beta_e}
\left\{ 
\begin{aligned}
 \beta_2^{(4- \varepsilon)} (u_{2}, u_{4}) &= - 2 u_2 - \left( 3 \pi + \cO(\varepsilon) \right)  u_{4} + \cO({u}^2)\,, \\
 \beta_4^{(4- \varepsilon)} (u_{2}, u_{4})  &= - \varepsilon u_4 - \left( 2 \pi + \cO(\varepsilon) \right) {u_{4}}^2 + \cO({u}^3)\,. 
\end{aligned}
\right.
\eeq

\

The flow equations (\ref{beta_e}) admit a non-trivial fixed point $(u_2^* , u_4^* )$, at a distance of order $\varepsilon$ away from the Gaussian fixed point:
\beq
\left\{ 
\begin{aligned}
u_2^* &= \frac{3}{4} \varepsilon + \cO(\varepsilon) \,, \\ 
u_4^* &= - \frac{1}{2 \pi} \varepsilon + \cO(\varepsilon) \,.
\end{aligned}
\right.
\eeq
This formal fixed point is very reminiscent of the Wilson--Fisher fixed point found in ordinary (Euclidean) quantum field theory in dimension $4 - \varepsilon$ \cite{wilson_fisher}. The main difference however is that, again due to an enhanced role of the wave-function renormalization in TGFT, the signs of the coupling constants are reversed with respect to the Wilson-Fisher fixed point. Hence $u_4^*$ has the 'wrong' sign as far as the convergence of the path-integral is concerned. Since we are ultimately only interested in the fate of this formal fixed point in the $\varepsilon \to 1$ limit, we will ignore this important aspect. 

\subsection{Properties of the non-Gaussian fixed point}

\begin{figure}[h]
  \centering
 	\includegraphics[scale=0.6]{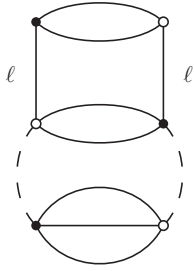}     
  \caption{The graph $G_3^\ell$ produces a one-loop quadratic contribution.}\label{g3l}
\end{figure}

In order to understand the behavior of the model in the vicinity of the non-trivial fixed point, we shall compute the linearized flow at $(u_2^* , u_4^*)$:
\beq
k\frac{\partial}{\partial k} \left( \begin{array}{c}   \delta u_2 \\  \delta u_4 \end{array} \right) = 
\left.\begin{pmatrix}
 \frac{\partial \beta_2}{\partial u_2} & \frac{\partial \beta_2}{\partial u_4} \\
 \frac{\partial \beta_4}{\partial u_2} & \frac{\partial \beta_4}{\partial u_4}
\end{pmatrix}
\right|_{(u_2^* , u_4^*)}  \left( \begin{array}{c}   \delta u_2 \\  \delta u_4 \end{array} \right)
+ \cO(\delta u^2)\,.
\eeq
If we want to compute the eigenvalues of this system up to order one in $\varepsilon$, it appears that the quadratic contributions to $\beta_2$ are needed. At one-loop we therefore need to include: a term proportional to $u_{2,k} u_{4,k}$ produced by the wave-function correction from the graphs $G_{1}^\ell$ (Figure \ref{g1l}); and a term proportional to $u_{2,k} u_{4,k}$ due to the graphs $G_{3}^\ell$ (see Figure \ref{g3l}).
This yields the following expression for the one-loop $\beta_2$:
\beq
\beta_2^{(4- \varepsilon)} (u_{2,k}, u_{4,k}) = - 2 u_{2,k} + 3 k^{\varepsilon - 1} \frac{\partial a_{k , \Lambda}^{(4- \varepsilon)}}{\partial k}  u_{4, k}  
+ \left( k^{1 + \varepsilon} \frac{\partial w_{k , \Lambda}^{(4- \varepsilon)}}{\partial k} - 3 k^{1 + \varepsilon} \frac{\partial b_{k , \Lambda}^{(4- \varepsilon)} }{\partial k} \right) u_{2,k} u_{4,k} + \cO({u_{k}}^3)\,, 
\eeq
and hence 
\beq
\beta_2^{(4- \varepsilon)} (u_{2}, u_{4}) = - 2 u_2 - 3 \pi \left( 1 - \varepsilon \ln \pi  + \cO(\varepsilon^2) \right)  u_{4} + \frac{3 \pi}{2}  \left( 1  + \cO(\varepsilon) \right)  u_2 u_{4}  + \cO({u}^3)\,.
\eeq
It is then easy to show that
\beq
\left.\begin{pmatrix}
 \frac{\partial \beta_2}{\partial u_2} & \frac{\partial \beta_2}{\partial u_4} \\
 \frac{\partial \beta_4}{\partial u_2} & \frac{\partial \beta_4}{\partial u_4}
\end{pmatrix}
\right|_{(u_2^* , u_4^*)}  = \begin{pmatrix}
 - 2 - \frac{3}{4} \varepsilon & - 3 \pi ( 1 - \frac{3}{8} \varepsilon - \varepsilon \ln \pi )   \\
 0 & \varepsilon
\end{pmatrix}
+ \cO(\varepsilon^2)
\eeq

At order $\varepsilon$ and at one-loop, the non-trivial fixed point $(u_2^* , u_4^*)$ has therefore one relevant eigendirection $V_r$ and one irrelevant eigendirection $V_{ir}$ with critical exponents $\theta_r$ and $\theta_{ir}$ respectively:
\beq
\left\{ \begin{aligned}
V_r &= \left( \begin{array}{c}   1 \\ 0 \end{array} \right)  \\
V_{ir} &= \left( \begin{array}{c} \frac{3 \pi}{2} ( 1 - \frac{5}{4} \varepsilon - \varepsilon \ln \pi ) \\ - 1 \end{array} \right)
\end{aligned}
\right.
\qquad {\rm{and}}
\qquad \left\{ 
\begin{aligned}
\theta_r &= 2 + \frac{3}{4} \varepsilon \\
\theta_{ir} &= - \varepsilon
\end{aligned}
\right.
\eeq

\

\begin{figure}[h]
  \centering
 	\includegraphics[scale=1]{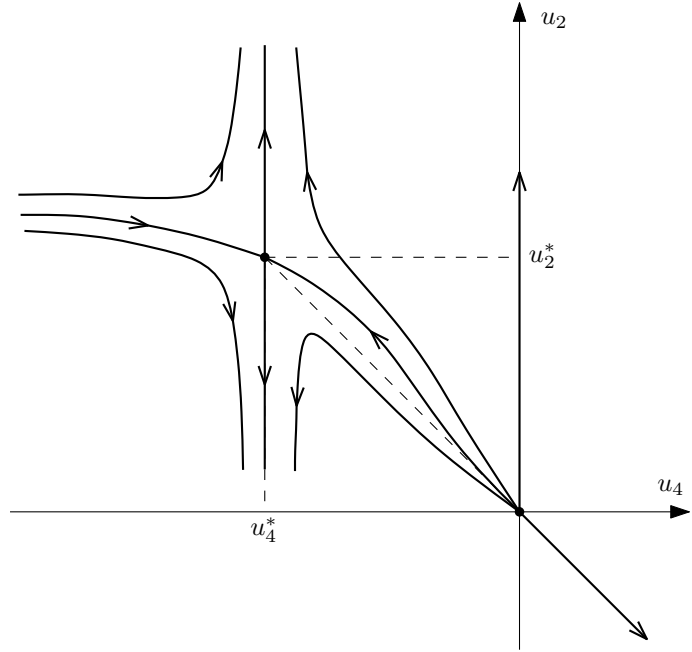}     
  \caption{Qualitative phase portrait for $D = 4 - \varepsilon$.}\label{portrait}
\end{figure}

We deduce the qualitative features of the phase portrait of this theory, represented in Figure \ref{portrait}. There exists a critical trajectory interpolating between the Gaussian fixed point and the non-Gaussian one. However the reader should be reminded that the equations we used to determine this fixed point are valid only in the deep ultraviolet sector. Hence one should expect a trajectory starting at $(u_2^* , u_4^*)$ in the UV to also start moving when reaching small enough values of the scale $k$. This is a manifestation of the non-autonomous nature of the exact flow, as also recently observed in \cite{frgft}. 
A second consequence of the presence of the non-trivial fixed point is a focusing of the trajectories in the $u_2$ direction towards the infrared. This suggests a scenario in which the theory generically becomes massive in the infrared, and therefore the Laplace operators in the kinetic term become negligible in this regime . This is particularly interesting in view of the question of the gravitational interpretation of such derivative operators in GFT, which might simply be evaded in the physical, small scale limit.

%% file: sections/su2.tex
We finally revisit, in the present language of the effective average action, the qualitative results about the $\SU(2)$ model already obtained in the discrete context \cite{discrete_rg}. This will allow us to speculate on the existence of a non-trivial fixed point also in this group dimension.

\subsection{One-loop beta functions}

\begin{figure}[h]
  \centering
 	\includegraphics[scale=0.6]{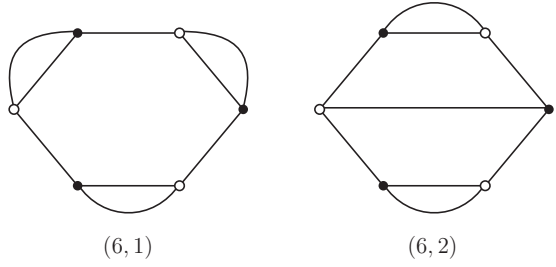}     
  \caption{Bubble interactions contributing to $S_{6,1}$ (left) and $S_{6,2}$ (right).}\label{s6}
\end{figure}

In $D=3$, two new $\vphi^6$ coupling constants become relevant, and the $\vphi^4$ interactions are super-renormalizable of dimension one. The effective average action is parametrized as follows:
\begin{align}
\Gamma_k (\vphib , \vphi) &= Z_k  k^2 u_{2,k}   S_2 (\vphib , \vphi) + Z_k S_\vphi (\vphib , \vphi) + {Z_k}^2 k \frac{u_{4 , k}}{2} S_4 (\vphib , \vphi) \\ \nn
\qquad &+ {Z_k}^3  \frac{u_{6,1 , k}}{3} S_{6,1} (\vphib , \vphi) + {Z_k}^3  u_{6,2 , k} S_{6,2} (\vphib , \vphi) + \cR_{\Lambda}^k (\vphib , \vphi) \,,
\end{align}
where $S_{6,1}$ and $S_{6,2}$ are defined by
\begin{align}
S_{6,1} (\vphi , \vphib) &=  \int [\extd g ]^9 \, \vphi(g_1 , g_2 , g_7 ) \vphib(g_1 , g_2 , g_9 ) \vphi(g_3 , g_4 , g_9 ) \vphib(g_3 , g_4 , g_8 ) \vphi(g_5 , g_6 , g_8 ) \vphib(g_5 , g_6 , g_7 ) \\
& + \; {\rm color} \; {\rm permutations}  \,, \nn \\
S_{6, 2} (\vphi , \vphib) &= \int [\extd g ]^9 \, \vphi(g_1 , g_2 , g_3 ) \vphib(g_1 , g_2 , g_4 ) \vphi(g_8 , g_9 , g_4 ) \vphib(g_7 , g_9 , g_3 ) \vphi(g_7 , g_5 , g_6 ) \vphib(g_8 , g_5 , g_6 ) \nn \\
& + \; {\rm color} \; {\rm permutations} \,.   
\end{align}
The two types of combinatorial structures contributing at order $6$ are represented in Figure \ref{s6}. 

\

Similarly to what has been done in \cite{discrete_rg}, we can compute the linear contributions to $\beta_2$ and $\beta_4$, but retain up to quadratic terms for the marginal coupling constants. In contrast with the discrete approach developed in \cite{discrete_rg} (to which we refer the reader for detailed computations) we moreover restrict to one-loop contributions, which simplifies further the analysis. One obtains:
\begin{align}
\beta_2  (u_{k})&= - 2 u_{2,k} + 3 \frac{\partial a_{k , \Lambda}^{(3)}}{\partial k}  u_{4, k} + \cO({u_{k}}^2)\,, \\ 
 \beta_4 (u_{k}) &=  -  u_{4,k} + 2 \frac{\partial a_{ k , \Lambda}^{(3)}}{\partial k}  u_{6,1, k} + 
 2 \frac{\partial a_{ k , \Lambda}^{(3)}}{\partial k}  u_{6,2, k} + \cO({u_{k}}^2)\,, \\
 \beta_{6,1} (u_{k}) &=  3 k^2 \left( \frac{\partial w_{ k , \Lambda}^{(3)}}{\partial k} - \frac{\partial b_{ k , \Lambda}^{(3)}}{\partial k} \right)  u_{4,k} u_{6,1, k} + \cO({u_{k}}^3 ) \,,\\
 \beta_{6,2} (u_{k}) &= k^2 \left( 3 \frac{\partial w_{ k , \Lambda}^{(3)}}{\partial k} - 2 \frac{\partial b_{ k , \Lambda}^{(3)}}{\partial k} \right)  u_{4,k} u_{6,2, k} + \cO({u_{k}}^3 )\,.
 \end{align} 
The numerical evaluation of the coefficients in the regime $1 \ll k \ll \Lambda$ gives:
\beq\label{flow_su2}
\left\{ 
\begin{aligned}
\beta_2  (u)&= - 2 u_{2} - 3 \sqrt{2 \pi} u_{4} + \cO({u}^2) \approx - 2 \, u_{2} - 7.5 \, u_{4} \,, \\ 
 \beta_4 (u) &=  -  u_{4} - 2 \sqrt{2 \pi}  u_{6,1} - 4 \sqrt{2 \pi}  u_{6,2} + \cO({u}^2) \approx -  u_{4} - 5.0  \, u_{6,1} - 10.0  \, u_{6,2} \,, \\
 \beta_{6,1} (u) &=  - 3 \left( 4 - \frac{5}{\sqrt{2}} \right)  u_{4} u_{6,1} + \cO({u}^3 ) \approx - 1.4 \, u_{4} u_{6,1} \,,\\
 \beta_{6,2} (u) &= - \left( 8 -\frac{7}{\sqrt{2}}\right)  u_{4} u_{6,1} + \cO({u}^3 ) \approx - 3.1 \, u_{4} u_{6,2}\,.
 \end{aligned}
 \right.
 \eeq 

\subsection{Phase portrait}



In addition to the Gaussian fixed point, the flow (\ref{flow_su2}) admits non-trivial fixed points with non-zero $u_{6,1}$ and $u_{6,2}$ of opposite signs, and $u_2 = u_4 = 0$. However, it is natural to expect this is nothing but a manifestation of our crude truncation. And indeed they do not survive the inclusion of two-loop graphs, as can for instance be seen in the discrete \cite{discrete_rg}.

\begin{figure}[h]
  \centering
 	\includegraphics[scale=0.6]{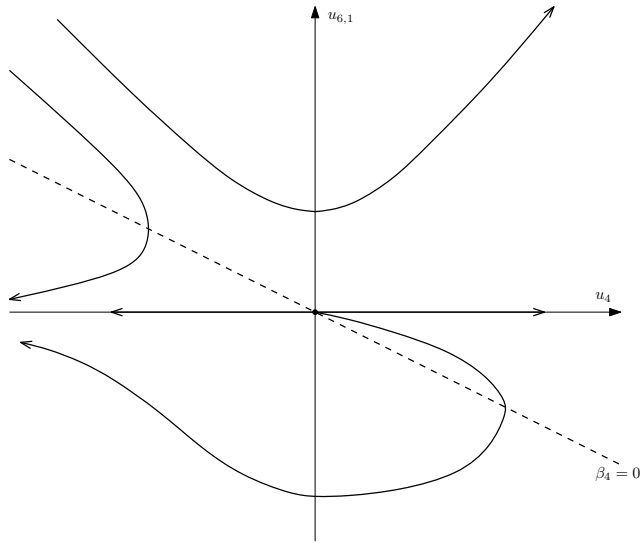}     
  \caption{Qualitative phase portrait for $D = 3$, in the stable $u_{6,2} = 0$ plane: we represented five typical trajectories, with arrows pointing towards the infrared.}\label{portrait2d}
\end{figure}

The perturbative result (\ref{flow_su2}) is only good for a qualitative understanding of the renormalization group flow in the vicinity of the Gaussian fixed point. One can for instance proceed along the lines of \cite{discrete_rg}, and wonder whether the theory becomes free in the deep UV. To answer this question we can discard the flow of the mass, which still leaves us with a three-dimensional and non-linear system. It is therefore convenient to exploit the fact that the hyperplanes $\left\{ u_{6,1} = 0 \right\}$ and $\left\{ u_{6,2} = 0 \right\}$ are invariant under this flow, and look at two-dimensional reductions of the latter. For example, one can prove that the phase portrait in the $\left\{ u_{6,2} = 0 \right\}$ plane is as represented in Figure \ref{portrait2d}, and similarly for $\left\{ u_{6,1} = 0 \right\}$. In this sector, there is no trajectory with $u_{6,1}>0$ which is at the same time asymptotically free: they can approach the Gaussian fixed point arbitrarily close, but are ultimately repelled in the $k \to + \infty$ limit. More generally, one can prove for the full system that there exists no asymptotically free trajectory with $u_{6,1} \geq 0$, $u_{6,2} \geq 0$ and at least one of them non-zero.

There still remains the possibility that trajectories with $u_{6,1}$ and $u_{6,2}$ of opposite signs could be made asymptotically free without ruining the convergence of the path-integral, as already noted in \cite{discrete_rg}. But this is a delicate question that we do not wish to address here. 

For $u_{6,1} > 0$ and $u_{6,2} > 0$ the partition function is well-defined. But can such a theory be UV completed? It could if a non-trivial fixed point were to prevent the divergence of the trajectories in the limit $k \to + \infty$. We cannot explore this further with perturbative techniques, but let us however remark that the non-trivial fixed point in $D = 4 - \varepsilon$ has a negative $u_4$, which is naively what one would look for in view of the phase portrait of Figure \ref{portrait2d} (and its three-dimensional analogue). 

Another rather intriguing possibility if a fixed point qualitatively similar to the one found in $D= 4 - \varepsilon$ is proven to exist also in $D = 3$ has to do with the bare propagator. The latter contains Laplace operators, but this is exclusively motivated by field theory: without the Laplace operators no consistent renormalization scheme can be implemented, as was first remarked in \cite{Valentin_Joseph}, in a slightly different but nonetheless very similar context. There is at this stage no satisfying gravitational understanding of such terms, which is the main reason why this $\SU(2)$ model was originally not proposed as a 3d quantum gravity model, but only as an interesting toy-model. The hypothetical fixed point might possibly change this situation because its presence could favor massive theories in the IR, as was already noted in $D= 4 - \varepsilon$. If so, propagators with derivative terms could be satisfactorily viewed as a property of the UV completed GFT, with no physical effect in the effectively continuous region which we expect to correspond to a small spin regime (with very many individual building blocks). There the propagator would become effectively ultra-local, and hence easily interpretable in terms of gravity spin foam amplitudes. This would henceforth completely evade the question of the gravitational origin of such derivative operators.

%% file: sections/conclu.tex





Let us briefly summarize our findings. Our primary goal was to introduce a $\varepsilon$--expansion in the context of 3d TGFTs with gauge invariance conditions. Along the way, we noticed important similarities with ordinary Euclidean scalar field theories. In $D = 4$ one obtains a renormalizable model with a single dimensionless $\vphi^4$ coupling
constant. The coupling constant acquires a non-zero
dimension when $D < 4$, and an analytic continuation allows to interpolate between $D = 3$ and
$D = 4$. For $D = 4 - \varepsilon$ and $\varepsilon$ small, a non-trivial fixed point is generated in the perturbative region, suggesting a qualitatively similar behavior when $\varepsilon =1$. However, in contrast to conventional quantum field theories, the $\vphi^4$ TGFT in $D=4$ is \emph{asymptotically free}, and for the same reason the coupling constants at the non--trivial fixed point in $D = 4 - \varepsilon$ have opposite signs with respect to the Wilson--Fisher fixed point. Also, the $\vphi^6$ model in $D=3$ has two marginal coupling constants rather than one, which makes the analysis of the flow more complicated.

\

The suggestion that the non-trivial fixed point in $D= 4 - \varepsilon$ might survive in the $\SU(2)$ model (or $U(1)^{\times 3}$) deserves to be explored. It represents a good opportunity for further applications of the FRG to GFTs, which would be undeniably valuable if they were available in 4d quantum gravity models. The FRG has recently been applied to matrix models and TGFTs without gauge invariance condition \cite{astrid_tim, astrid_tim_II, frgft}, hence we think that the $D=3$ model of the present paper is the natural next step in this research direction, and it could provide a UV completion of otherwise diverging renormalization group trajectories. We also proposed a speculative idea towards a more physical application, would the existence of this fixed point be confirmed: its presence might favor dynamically generated ultra-local theories in the infrared, hence possibly evading the question of the gravitational interpretation of the bare Laplace-type propagator.